\documentclass[prd,twocolumn,amssymb,amsmath]{revtex4}

\usepackage{graphicx}
\usepackage{txfonts}
\usepackage{aas_macros}

\usepackage{color}

\newcommand{\beq}{\begin{equation}}   %

\newcommand{\eeq}{\end{equation}}   %

\newcommand{\beqa}{\begin{eqnarray}}   %

\newcommand{\eeqa}{\end{eqnarray}}   %

\newcommand{\beal}{\begin{align}}

\newcommand{\enal}{\end{align}}

\newcommand{\bspl}{\begin{split}}

\newcommand{\espl}{\end{split}}

\newcommand{\bsub}{\begin{subequations}}

\newcommand{\esub}{\end{subequations}}

\newcommand{\bmulti}{\begin{multline}}   %

\newcommand{\beqm}{\begin{mathletters}}   %

\newcommand{\eeqm}{\end{mathletters}}   %

\begin{document}
\newcommand{\bx}{{\bf x}}
\newcommand{\bn}{{\bf n}}
\newcommand{\bk}{{\bf k}}
\newcommand{\dd}{{\rm d}}
\newcommand{\dslash}{D\!\!\!\!/}
\def\ga{\mathrel{\raise.3ex\hbox{$>$\kern-.75em\lower1ex\hbox{$\sim$}}}}
\def\la{\mathrel{\raise.3ex\hbox{$<$\kern-.75em\lower1ex\hbox{$\sim$}}}}
\def\beq{\begin{equation}}
\def\eeq{\end{equation}}

\vskip-2cm
\title{\textcolor{black}{$\Delta N_{\text{eff}}$ and entropy production from early-decaying gravitinos}}

\author{{Emanuela Dimastrogiovanni$^a$$^,$$^b$ and Lawrence M. Krauss$^b$$^,$$^c$}}
\affiliation{
   $^a$ Department of Physics/CERCA/Institute for the Science of Origins,
Case Western Reserve University, Cleveland, OH 44106, USA\\
     $^b$Department of Physics and School of Earth and Space Exploration, Arizona State University, Tempe, AZ 85827, USA\\
     $^c$Research School of Astronomy and Astrophysics, Mt. Stromlo Observatory, Australian National University,\\ Canberra, Australia 2611  }

\vspace*{1cm}

\begin{abstract}

Gravitinos are a fundamental prediction of supergravity, their mass ($m_{G}$) is informative of the value of the SUSY breaking scale, and, if produced during reheating, their number density is a function of the reheating temperature ($T_{\text{rh}}$). As a result, constraining their parameter space provides in turn significant constraints on particles physics and cosmology. We have previously shown that for gravitinos decaying into photons or charged particles during the ($\mu$ and $y$) distortion eras, upcoming CMB spectral distortions bounds are highly effective in constraining the $T_{\text{rh}}-m_{G}$ space. For heavier gravitinos (with lifetimes shorter than a few $\times10^6$ sec), distortions are quickly thermalized and energy injections cause a temperature rise for the CMB bath. If the decay occurs after neutrino decoupling, its overall effect is a suppression of the effective number of relativistic degrees of freedom ($N_{\text{eff}}$). In this paper, we utilize the observational bounds on $N_{\text{eff}}$ to constrain gravitino decays, and hence provide new constaints on gravitinos and reheating.  For gravitino masses less than $\approx 10^5$ GeV, current observations give an upper limit on the reheating scale in the range of $\approx 5 \times 10^{10}- 5 \times 10^{11}$GeV.  For masses greater than $\approx 4 \times 10^3$ GeV they are more stringent than previous bounds from BBN constraints, coming from photodissociation of deuterium, by almost 2 orders of magnitude.

\end{abstract}

\date{\today}

\maketitle



\section{Introduction} 
\label{introduction}

The gravitino is predicted in supergravity as the spin $3/2$ superpartner of the graviton (see e.g. \cite{reviewsg} for a review). If supersymmetry is broken the gravitino gets a mass determined by the supersymmetry breaking scale: $F\approx\sqrt{m_{G}\,M_{P}}$, $m_{G}$ being the gravitino mass and $M_{P}\approx 2.4\times 10^{18}\text{GeV}$. While gravitinos would be thermally produced in the early universe they would be diluted by the expansion during Inflation. Following Inflation, they would again be produced during reheating.  There is thus a direct relation between the reheating temperature, $T_{\text{rh}}$, and the gravitino number density. The importance of constraining the reheating temperature cannot be overstated: $T_{\text{rh}}$ is one of handful of macroscopic parameters  describing the transition from an early phase of accelerated expansion (\textsl{inflation}) to the radiation-dominated era and it sets a lower bound on the energy scale of inflation ( e.g. \cite{reheating}).\\

Because of their potentially large masses, gravitinos can have significant cosmological impacts \cite{reviewco}. An overabundance of long-lived gravitinos (or the decay products of unstable ones) may overclose the universe. Gravitinos decaying during or after big-bang nucleosynthesis (BBN) may alter the expansion rate of the universe, produce a suppression of the baryon-to-photon ratio and, most importantly, destroy the light elements, thus altering the successful predictions of standard BBN \cite{standardbbn}. In addition, gravitinos decaying into photons or baryons after the onset of the $\mu$-distortion era ($z\lesssim 2\times 10^6$), would generate distortions of the cosmic microwave background (CMB) black-body spectrum. This was discussed in \cite{previousSDS}, and more recently in \cite{Dimastrogiovanni:2015wvk}. The sensitivity limits on the $\left(T_{\text{rh}},\,m_{G}\right)$ parameter space for a PIXIE-like experiment  for upcoming spectral distortions probes will be much more stringent than current BBN bounds. \\

The goal of this work is to complement the analysis in \cite{Dimastrogiovanni:2015wvk} for gravitino produced during the reheating era by investigating the parameter space relevant for earlier ($z\gtrsim 2\times 10^6$) decays. At those high redshifts, thermalization processes in the hot plasma are highly efficient and quickly erase any produced distortions. The net effect of an energy injection from gravitino decay in the primordial bath would be a temperature increase for the CMB, as well as for all relativistic species coupled to it. \\
\indent In the minimal cosmological scenario, the total radiation energy density after electron/positron annihilation ($T\lesssim T_{e^{\pm}}\simeq 0.5\,\text{MeV}$) is contributed by photons and neutrinos, and parametrized as
\begin{equation}\label{eq.1a}
\rho_{R}\approx \frac{\pi^2}{15}\left[1+\frac{7}{8}N_{\nu}\left(\frac{T_{\nu}}{T}\right)^{4}\right] T^4\,.
\end{equation}
Here $N_{\nu}$ is the number of neutrinos species and $T_{\nu}$ their temperature. Neutrinos remain in thermal equilibrium with the CMB until their interaction rate with other standard model particles drops below the expansion rate (at $T\approx 1\,\text{MeV}$). After decoupling, neutrino temperature remains approximately equal to the CMB temperature until $T=T_{e^{\pm}}\approx 0.5\,\text{MeV}$: around this time the entropy released from electron/positron annihilation causes the CMB temperature to rise, while leaving the neutrino temperature nearly unaffected \footnote{Because of the proximity of the time of neutrino decoupling to $T_{e^{\pm}}$, neutrinos are mildly heated by $e^{+}e^{-}$ annihilations. This produces small non-thermal distortions in the neutrino spectra and eventually results into a slightly higher value for the neutrino-to-photon temperature ratio (or, equivalently, for the effective number of neutrino species) than one obtains in the instantaneous decoupling approximation \cite{Fornengo:1997wa}.}. Assuming instantaneous neutrino decoupling, this yields  $T_{\nu}/T= (4/11)^{1/3}$. In a similar way, a decay of gravitinos into photons or baryons, taking place between neutrino decoupling and the onset of the distortion eras, would result in an increase of $T/T_{\nu}$. Eq.~(\ref{eq.1a}) is often rewritten, in more general terms, as
\begin{equation}
\rho_{R}\approx  \frac{\pi^2}{15}\left[1+\frac{7}{8} \left(\frac{4}{11}\right)^{4/3} N_{\text{eff}}\right] T^4\,,
\end{equation}
with the parameter $N_{\text{eff}}\equiv (1/4)^{4/3} N_{\nu}(T_{\nu}/T)^{4}$ quantifying the effective number of non-photonic relativistic degrees of freedom.\\
\indent In the standard model of particle physics, with three active neutrino species, $N_{\nu}$ is slightly larger than 3 if one accounts for relic interactions between electrons and neutrinos during the time of $e^{\pm}$ annihilation. The resulting value is $N_{\nu}=3.046$, which also incorporates finite temperature QED corrections to the electromagnetic plasma and flavour oscillations effects \cite{Fornengo:1997wa}.\\

$N_{\text{eff}}$ is constrained in a number of ways: by the predictions of BBN, paired with observations of light elements abundances \cite{bbnle}; by CMB temperature and polarization anisotropies \cite{cmbne}; by the large scale structure (LSS) of the matter distribution \cite{lssne}. Within current experimental bounds, all of the aforementioned probes show agreement with the standard prediction of $N_{\text{eff}}=3.046$ \cite{currentne}. On the other hand, current limits allow ample room for deviations (a non-zero $\Delta N_{\text{eff}}\equiv N_{\text{eff}}-3.046$) which would signal new physics. Future observations are expected to greatly improve on the present bounds (see e.g. \cite{futurene}).\\

Additional radiation density from non-standard-model degrees of freedom may result in $\Delta N_{\text{eff}}>0$ \cite{axions,sterileneutrinos,asm,Boehm:2012gr,Kanzaki:2007pd,Hooper:2011aj,Mangano:2006ar}.  On the other hand, the scenario we describe here, where additional photons or charged particles are produced after neutrino decoupling, results instead in a suppression of $N_{\text{eff}}$. (Note that other proposals predicting $\Delta N_{\text{eff}}<0$ include models in which the neutrino thermalization remains incomplete, as in low-reheating models \cite{deSalas:2015glj}). \\

This work is organized as follows: in Sec.~\ref{overview} we briefly review results for the thermal production of gravitinos during reheating and we formally introduce the relation between $\Delta N_{\text{eff}}$ and model parameters; in Sec.~\ref{entropy} we compute $\Delta N_{\text{eff}}$ from gravitino decays and place constraints on the reheating temperature and gravitino mass parameter space; in Sec.~\ref{conclusions} we comment on implications for gravitinos of future constraints on $N_{\text{eff}}$ arising from measurements of CMB anisotropies and LSS observations.


\section{Gravitino decays and cosmology} 
\label{overview}

For gravitinos produced (thermally) from interactions in the hot plasma during reheating, there is a simple relation between number density and the temperature at the end of reheating \cite{Ellis:1984eq}:
\begin{equation}\label{grav1}
n_{G}
=Y_{G}\,s(T),\quad\quad Y_{G}\approx 10^{-12}\frac{T_{\text{rh}}}{10^{10}\text{GeV}}\,,
\end{equation}
where $s(T)\equiv (2\pi^2/45)g_{*}(T)T^3$ is the entropy density and $g_{*}$ is the number of relativistic degrees of freedom.\\
The gravitinos decay rate can be parametrized as follows \cite{Krauss:1983ik} 

\begin{equation}\label{two}
\Gamma_{G}=\frac{N_{\text{dec}}}{2\pi}\frac{m_{G}^{3}}{M_{P}^{2}}\,,
\end{equation}
where $N_{\text{dec}}$ is the effective number of decay channels. For gravitinos decaying into photons and photinos, $G\rightarrow \gamma+\tilde{\gamma}$, and for negligible photino mass ($m_{\tilde{\gamma}}\ll m_{G}$), one finds $N_{\text{dec}}\approx 1/16$. For decay into hadrons $N_{\text{dec}}\approx 2/5$. \\

\noindent The total radiation energy density after BBN is given by  
\begin{eqnarray}\label{eq.1}
\rho_{R}&=&\sum_{i}\rho_{i}=\frac{\pi^2}{30} \left[\sum_{i=\text{bosons}}g_{i}T_{i}^{4}+\frac{7}{8}\sum_{i=\text{fermions}}g_{i}T_{i}^{4}\right]\nonumber\\&=& \frac{\pi^2}{30}g_{*}(T)T^4\,,
\end{eqnarray}
where $T$ is the CMB temperature and
\begin{equation}\label{energy1}
g_{*}(T)\equiv  \sum_{i=\text{bosons}}g_{i}\left(\frac{T_{i}}{T}\right)^{4}+\frac{7}{8}\sum_{i=\text{fermions}}g_{i}\left(\frac{T_{i}}{T}\right)^{4}\,.
\end{equation}
If gravitinos decay into particles heating up the CMB and the decay occurs before the onset of the $\mu$ era ($z\gtrsim 2\times 10^6$ or $t\lesssim 6\times 10^6\,\text{sec.}$) and after electron decoupling, the resulting temperature increase for the CMB w.r.t. the neutrinos temperature leads to a smaller value for $N_{\text{eff}}$ than one would observe in the absence of those decays. In the simplified case where the decay happens instantaneously at $t=t_{G}$, one expects 
\begin{equation}
N_{\text{eff}}\propto \begin{cases}
    N_{\nu}\,, & \text{if $\,t<t_{G}$}.\\
     N_{\nu}\,f(m_{G},T_{\text{rh}},N_{\text{dec}})\,, & \text{if $\,t > t_{G}$}\,.
  \end{cases}
\end{equation}
Here, $f(m_{G},T_{\text{rh}},N_{\text{dec}})$ parameterizes the impact of gravitino decays and is derived in the next section.


\section{Entropy injection and $\Delta N_{eff}$ constraints} 
\label{entropy}

For instantaneous decay and thermalization of the decay products, energy conservation right before and after the decay implies \footnote{Notice that it is only necessary to account for radiation and for gravitinos in the energy balance: if other massive particles were present, but not undergoing decay, they would add up to both sides of Eq.~(\ref{one}). The same applies more generally to all species, such as neutrinos, that by time  $t_{G}$ have decoupled from the CMB.}:
\begin{eqnarray}\label{one}
\rho^{\text{before}}_{\text{total}}&=&\frac{\pi^2}{30}g_{*}^{\text{th}}(t_{G}) T^{4}(t_{G})+\frac{2\pi^2}{45}g_{*}^{\text{th}}(t_{G})T^{3}(t_{G}) \,m_{G}\,\epsilon_{G}\,Y_{G}\nonumber\\&=&\rho^{\text{after}}_{\text{total}}=\frac{\pi^2}{30}g_{*}^{\text{th}}(t') T^4(t')\,.
\end{eqnarray}
%
The $\epsilon_{G}$ parameter accounts for the actual fraction of the gravitinos energy density that after decay is transferred into the CMB (through Comptonization). For gravitinos decaying entirely into photons$+$photinos, the reasonable expectation is that roughly half of the initial energy would be converted into heating, hence $\epsilon_{G}\approx 1/2$. For decays into colored particles, one would expects much more efficient heating ($\epsilon_{G}\approx 1$). \\
Setting $g_{*}^{\text{th}}(t_{G})=g_{*}^{\text{th}}(t')$:
\begin{equation}
\left(\frac{T(t')}{T(t_{G})}\right)^{4}=1+\frac{4}{3}\frac{m_{G}\,\epsilon_{G}Y_{G}}{T(t_{G})}\frac{g_{*(s)}(t_{G})}{g_{*}^{\text{th}}(t_{G})}\,.
\end{equation}
We need to find $N_{\text{eff}}\propto (T(t'')/T_{\nu}(t''))^4$, for a generic time $t''>t'$. Let us require entropy conservation between $t'$ and $t''$
\begin{equation}
g_{*s}^{\text{th}}(t')a^{3}(t')T^{3}(t')=g_{*s}^{\text{th}}(t'')a^{3}(t'')T^{3}(t'')\,.
\end{equation}
Introducing the scaling law for the neutrino temperature
\begin{equation}
T_{\nu}(t'')=T_{\nu}(t_{\nu})\frac{a(t_{\nu})}{a(t'')}=T(t_{\nu})\frac{a(t_{\nu})}{a(t'')}\,,
\end{equation}
$t_{\nu}$ being the time of neutrino decoupling, and requiring entropy conservation between $t_{\nu}$ and $t_{G}$
\begin{equation}
g_{*s}^{\text{th}}(t_{\nu})a^{3}(t_{\nu})T^{3}(t_{\nu})=g_{*s}^{\text{th}}(t_{G})a^{3}(t_{G})T^{3}(t_{G})\,,
\end{equation}
one arrives at
\begin{equation}
\frac{T(t'')}{T_{\nu}(t'')}=\left(\frac{g_{*s}^{\text{th}}(t_{\nu})}{g_{*s}^{\text{th}}(t'')}\right)^{1/3}\left[1+\frac{4}{3}\frac{m_{G}\,\epsilon_{G}Y_{G}}{T(t_{G})}\left(\frac{g_{*s}(t_{G})}{g_{*}^{\text{th}}(t_{G})}\right)\right]^{1/4}\,.
\end{equation}

\begin{center}
\begin{figure}
\includegraphics[width=0.49\textwidth]{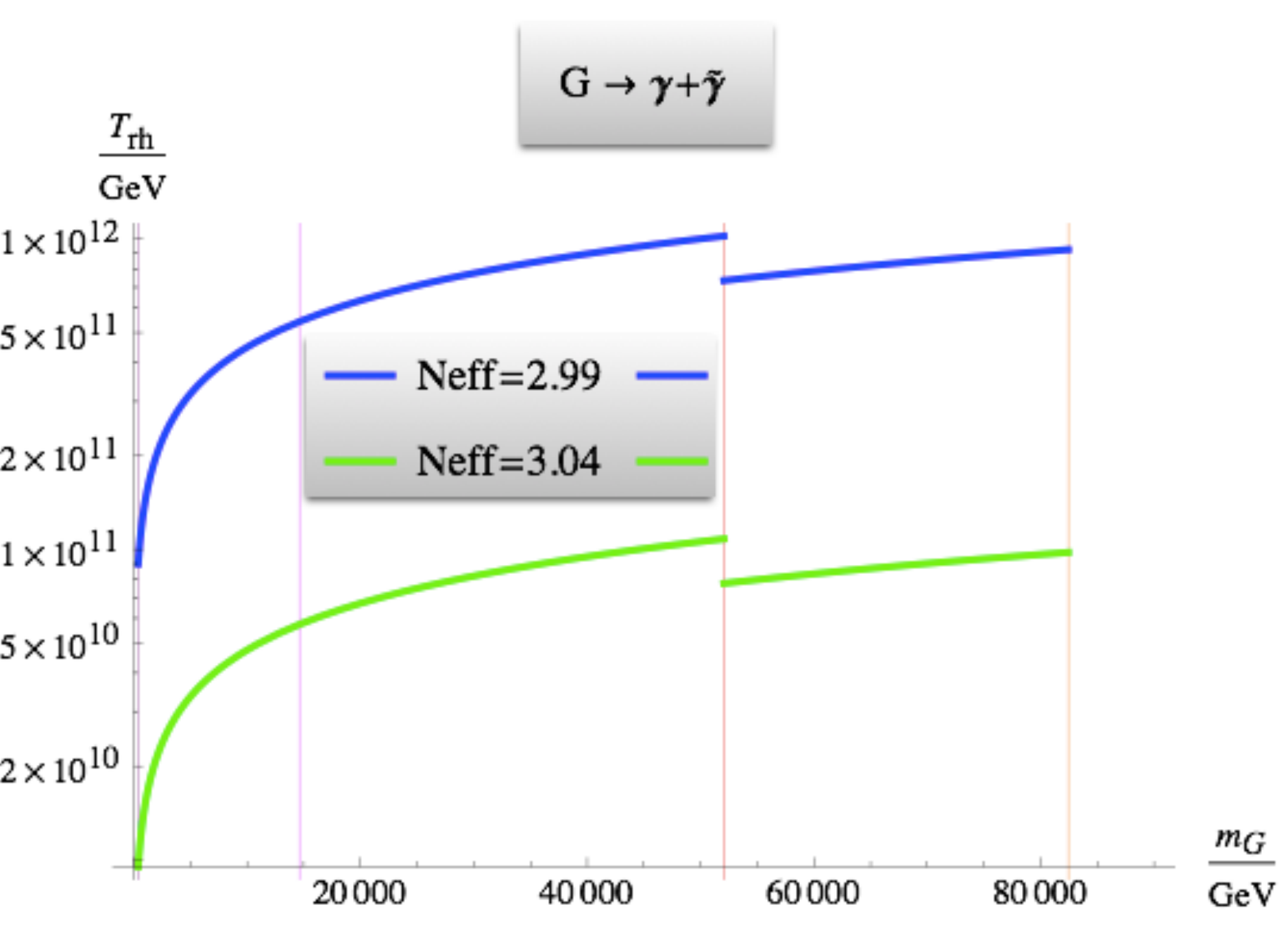}
\includegraphics[width=0.49\textwidth]{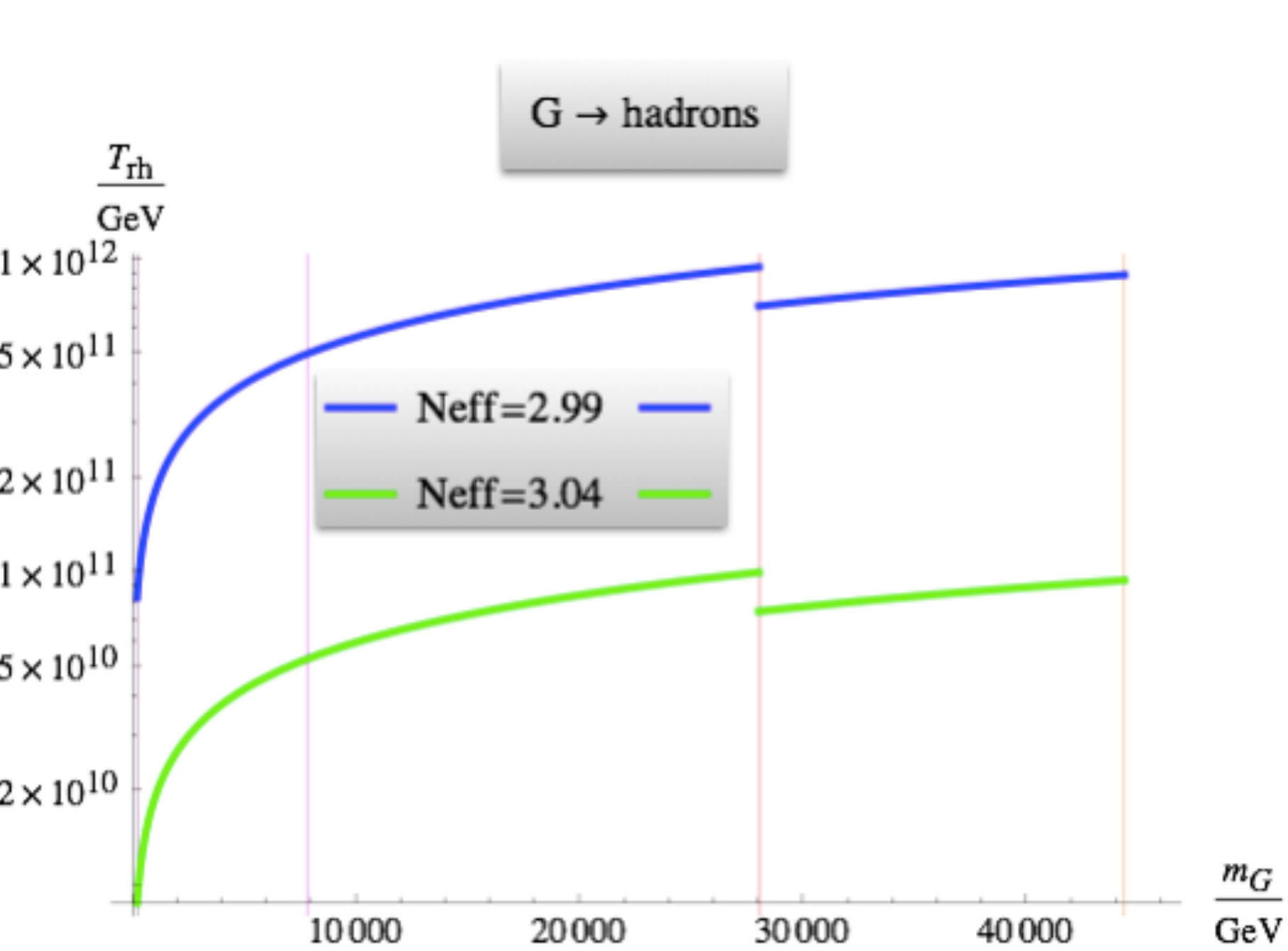}
  \caption{Upper panel: bounds on the reheating temperature for $N_{\text{eff}}=2.99$ (blue lines) and $N_{\text{eff}}=3.04$. We have chosen $N_{\nu}=3.046$. The vertical lines correspond to gravitino masses $m_{G}\in[8.2 \times 10^4, 5.2 \times 10^4, 1.5 \times 10^4, 4 \times 10^2]\,\text{GeV}$, i.e. decaying respectively around $t_{\nu}, t_{e^{\pm}}, t_{BBN}, t_{\mu}$. {The jump at gravitino mass around $5.2 \times 10^4$ GeV corresponds to the extra deposition of entropy into thermal electrons and positrons before they annihilate.} Lower panel: entropy production bounds on the reheating temperature for hadronic decays. In this case the gravitino masses corresponding to the times referred to above are reduced by $\approx 1.8$, and the jump occurs at $2.8 \times 10^4$ GeV.}
\label{fig1}
\end{figure}
\end{center}

The final expression for $N_{\text{eff}}$ is then given by
\begin{equation}\label{compp}
N_{\text{eff}}\simeq N_{\nu}\left[1+\frac{4}{3}\frac{m_{G}\,\epsilon_{G}Y_{G}}{T(t_{G})}\left(\frac{g_{*s}(t_{G})}{g_{*}^{\text{th}}(t_{G})}\right)\right]^{-1}\,.
\end{equation}
For a decay $G\rightarrow \gamma+\tilde{\gamma}$, one sets $N_{\text{dec}}=1/16$ (from Eq.~(\ref{two})) and $\epsilon_{G}=1/2$. Eq.~(\ref{compp}) can then be recast in the following form
\begin{eqnarray}\label{finen3}
N_{\text{eff}}=N_{\nu}\left[1+\tilde{\omega}\left(1+\frac{7}{22}N_{\nu}\right)\left(\frac{m_{G}}{\text{GeV}}\right)^{-1/2}\left(\frac{T_{\text{rh}}}{\text{GeV}}\right)\right]^{-4/3}\,,
\end{eqnarray}
where, taking account of the entropy transferred to the photon bath after $e^{\pm}$ annihilation, 
\begin{equation}
\begin{cases}
    \tilde{\omega}\approx  2.2\times 10^{12}  & \text{for}\quad t_{G}\lesssim t_{e^{\pm}}\,,\\
     \tilde{\omega}\approx  1.6\times 10^{12}  & \text{for}\quad t_{G}> t_{e^{\pm}}\,.
  \end{cases}
\end{equation}
One can then employ the known bounds on $N_{\text{eff}}$ and $N_{\nu}$ to constraints the $(T_{rh}, m_{G})$ parameter space. Making the conservative assumption that $N_{\nu}=3.046$ (i.e. ignoring the possibility of extra neutrino species, which would only serve to strengthen the constraints we derive here), Eq.~(\ref{finen3}) can be further simplified
\begin{equation}\label{finen7}
\frac{T_{\text{rh}}}{\text{GeV}}=\left[\left(\frac{N_{\nu}}{N_{\text{eff}}}\right)^{3/4}-1\right]\left(\frac{m_{G}}{\text{GeV}}\right)^{1/2}\tilde{\alpha}\,,
\end{equation}
where 
\begin{equation}
\begin{cases}
    \tilde{\alpha}\approx  2.3\times 10^{11}  & \text{for}\quad t_{G}\lesssim t_{e^{\pm}}\,,\\
     \tilde{\alpha}\approx  3.2\times 10^{11}  & \text{for}\quad t_{G}> t_{e^{\pm}}\,.\\
  \end{cases}
\end{equation}
For the temperature in Eq.~(\ref{finen7}) to be positive definite, the condition $N_{\text{eff}} < N_{\nu}$ must hold. Any given $N_{\text{eff}}<N_{\nu}$ defines a curve in the $(T_{rh}, m_{G})$ plane: the smaller the ratio $N_{\nu}/N_{\text{eff}}$, the smaller the value of the predicted reheating temperature (or, equivalently, of the gravitinos number density) for a given mass $m_{G}$. \\

As mentioned, constraints on $N_{\text{eff}}$ can be extracted from CMB anisotropies and LSS data. The number of neutrino species affects the value of the photon diffusion damping scale. In addition, it contributes to the total radiation density and therefore it impacts the time of matter-radiation equality and the expansion rate of the universe. This has consequences on the location and amplitude of the acoustic peaks (\textit{l}$\,\,\gtrsim200$ multipoles) in the temperature and polarization power spectra, and on the shape and overall amplitude of the cold dark matter power spectrum \cite{cmbne,lssne}.\\
\indent Current bounds on $N_{\text{eff}}$ from Planck (also in combination with data sets from baryon acoustic oscillation (BAO) measurements) are compatible with $N_{\text{eff}}<N_{\nu}$. We report here the $68\%$CL constraints from \cite{currentne} 
\begin{eqnarray} 
&&N_{\text{eff}}=3.13\pm0.32 \,\,\,\,\,\,[\text{\textit{Planck}TT}+\text{lowP}]\,,\\  
&&N_{\text{eff}}= 3.15\pm0.23 \,\,\,\,\,\,[\text{\textit{Planck}TT}+\text{lowP}+\text{BAO}]\,,\\  \label{pdata1}
&&N_{\text{eff}}= 2.99\pm0.20\,\,\,\,\,\,[\text{\textit{Planck}TT,TE,EE}+\text{lowP}]\,,\\   \label{pdata2}&&N_{\text{eff}}=3.04\pm0.18\,\,\,\,\,\,[\text{\textit{Planck}TT,TE,EE}+\text{lowP}+\text{BAO}]\,.\nonumber\\
\end{eqnarray}

In order to demonstrate the range of bounds possible from measurements of $N_{\text{eff}}$, the upper panel of Fig.~(\ref{fig1}) shows the lines corresponding to the central values in Eqs.~(\ref{pdata1}) and (\ref{pdata2}). A similar analysis for hadron decay leads to the constraints reported in the lower panel of Fig.~\ref{fig1}.\\
\indent From Eq.~(\ref{two}) one has $m_{G}=[(2\pi M_{P}^{2})/(t_{G} N_{\text{dec}})]^{1/3}$. The plots highlight the values for the gravitino mass that correspond to four benchmark values of (decay) time: neutrino decoupling ($t_{\nu}$); electron/positron annihilation ($t_{e^{\pm}}$); end of BBN ($t_{BBN}$); onset of $\mu$ distortion era ($t_{\mu}$).  {The discontinuities in the bound at a gravitino mass of 52 TeV for photon decay and at 28 TeV for hadronic decays reflect the changing relation between entropy dumped in the CMB and gravitino masses for gravitinos which decay before and after  $e^{\pm}$ annihilation.}\\

\indent As in \cite{Dimastrogiovanni:2015wvk}, it is useful to draw a comparison with other cosmological bounds on unstable gravitinos. These arise primarily from BBN predictions for light elements abundances (see e.g. \cite{Krauss:1983ika}). Unlike for $N_{\text{eff}}$, these constraints do not define an exact relation between $T_{\text{rh}}$ and $m_{G}$,  nor do they require a specific sign for $\Delta N_{\text{eff}}$, however they are able to rule out conspicuous portions of the parameter space.  BBN constraints are important in the lower end of the mass range of Fig.~(\ref{fig1}), as they generally require gravitinos to decay after deuterium production during standard BBN is complete. In this respect, the two probes may well be regarded as complementary to one another. {As an example, for decays into photons and photinos, BBN limits the reheating temperature to values below $10^{8}-10^{9}$ GeV for $4\times 10^{2}\lesssim m_{G}\lesssim  10^{3}\,\text{GeV}$. These would be nearly two orders of magnitude more stringent than the upper bounds on the reheating temperature in this mass range for $N_{\text{eff}}=2.99$. For $m_{G}\gtrsim 3\times 10^{3}\,\text{GeV}$, the bounds from BBN become weaker than those shown in Fig.~\ref{fig1}, by two orders of magnitude or more, moving towards heavier gravitinos.}    \\

\indent Because of the sensitivity we have demonstrated here of gravitino bounds to $N_{\text{eff}}$, a tightening of the bounds on $N_{\text{eff}}$ coming from upcoming CMB observations should allow significantly improved parameter space restrictions for post-inflation gravitino production and decay.


\section{Conclusions and outlook} 
\label{conclusions}

Entropy transfers to the CMB bath from (additional) heavy particles, decaying after neutrino decoupling, suppress the ratio of the neutrino-to-CMB temperatures thereafter or, in other words, the effective number of relativistic species, $N_{\text{eff}}$, w.r.t. its standard model prediction ($N_{\text{eff}}\simeq 3.046$). \\
\indent This would be the case for unstable gravitinos decaying in the pre-distortion eras. For gravitinos generated during reheating, simple relations hold between their number density and the reheating temperature, $T_{\text{rh}}$, making the constraints on their parameter space all the more interesting for cosmology. \\
\indent We have derived an analytic relation among the theory parameters ($T_{\text{rh}}$, the gravitino mass, $m_{G}$, and $N_{\text{dec}}$, describing the branching ratios of the decay) and $N_{\text{eff}}$. For a given set of decay channels, specifying a value of $N_{\text{eff}}< 3.046$ yields a specific $T_{\text{rh}}-m_{G}$ relation. This is presented in Fig.~(\ref{fig1}), both for photon and hadron decays, and for selected measured values for $N_{\text{eff}}$ from the Planck combined analysis of temperature and polarization anisotropy ($+$ BAO) data. \\
\indent The mass range analyzed in this work is complementary both to the one that can be probed with spectral distortions \cite{Dimastrogiovanni:2015wvk} (e.g. one needs $m_{G}\lesssim 700\,\text{GeV}$ for gravitinos decays into photons$+$photinos to produce $\mu$ or $y$ distortions) and to the one constrained from BBN (the latter being more effective towards the lower end of our mass range). \\
\indent {For lighter gravitinos ($m_{G}\lesssim  10^{3}\,\text{GeV}$), the BBN bounds on the reheating temperature are nearly 2 order of magnitude stronger than those given by a value $N_{\text{eff}}=2.99$ (corresponding to the blue lines in Fig.~\ref{fig1}), whereas the situation is reversed for heavier masses. In the range $3\times 10^{3}\lesssim m_{G}\lesssim  10^{5}\,\text{GeV}$, for example, $N_{\text{eff}}=2.99$ would constrain the reheating temperature to values $\lesssim 5\times 10^{11}$ GeV.} \\
 
Future CMB and LSS observations hold great promise for further constraining $N_{\text{eff}}$ and could rule out or confirm with very high significance the standard model value, providing sensitivity to new physics. These include the next generation ground-based CMB experiments (S4), with a sensitivity forecast of $\sigma(N_{\text{eff}})\sim 0.03$, and the proposed CORE space mission, which would reach $\sigma(N_{\text{eff}})\sim 0.04 $ \cite{futurene} (also in combination with future data from galaxy surveys such as DESI \cite{http://desi.lbl.gov.} and Euclid \cite{euclid}). This will have enormous implications for our ability to test physics beyond-the-standard model as well as neutrino physics, and possibly resolve some of the apparent discrepancies in cosmological data (including, e.g., the tension between $H_{0}$ direct measurements and its estimates from CMB observations). Models predicting a suppression of $N_{\text{eff}}$, of the kind considered in this paper, could be ruled out by these future experiments. Alternatively, another interesting phenomenological implication of these early-decaying gravitino scenarios is that, by reducing the contribution to $N_{\text{eff}}$ from neutrinos (being $T_{\nu}/T$ lower than in the standard case), they leave more room for positive contributions to $\Delta N_{\text{eff}}$ from the dark-sector.


\section*{Acknowledgments}

{E.D. and L.M.K. acknowledge support from the Department of Energy under grant No. DE-SC0008016. E.D. was also supported by DOE grant DE-SC0009946.}


\end{document}